\UseRawInputEncoding
\documentclass[aps, prl, preprint, graphics, showpacs, amssymb, amsmath, superscriptaddress]{revtex4-1}
\usepackage{graphicx}
\usepackage{dcolumn}
\usepackage{bm}
\usepackage{epstopdf}
\usepackage{amssymb}
\usepackage{color}
\usepackage[colorlinks=true, pdfstartview=FitV, linkcolor=blue, citecolor=blue, urlcolor=blue]{hyperref}
\usepackage{lineno}

\usepackage{soul}
\begin{document}

\title{SeWS/bilayer-SiC heterojunction: An S-scheme photocatalyst with high visible-light absorption, excellent carrier mobility and adjustable band gap}
\author{Liuzhu Yang}
\affiliation{State Key Laboratory of Metastable Materials Science and Technology \& Hebei Key Laboratory of Microstructural Material Physics, School of Science, Yanshan University, Qinhuangdao 066004, China }
\author{Wenhui Wan}
\affiliation{State Key Laboratory of Metastable Materials Science and Technology \& Hebei Key Laboratory of Microstructural Material Physics, School of Science, Yanshan University, Qinhuangdao 066004, China }
\author{Zhicui Wang}
\affiliation{Qinhuangdao Vocational and Technical College, Qinhuangdao 066100, China}
\author{Qiuyue Ma}
\affiliation{State Key Laboratory of Metastable Materials Science and Technology \& Hebei Key Laboratory of Microstructural Material Physics, School of Science, Yanshan University, Qinhuangdao 066004, China }
\author{Yanfeng Ge}
\affiliation{State Key Laboratory of Metastable Materials Science and Technology \& Hebei Key Laboratory of Microstructural Material Physics, School of Science, Yanshan University, Qinhuangdao 066004, China }
\author{Yong Liu}\email{yongliu@ysu.edu.cn }
\affiliation{State Key Laboratory of Metastable Materials Science and Technology \& Hebei Key Laboratory of Microstructural Material Physics, School of Science, Yanshan University, Qinhuangdao 066004, China }

\begin{abstract}
Vertically stacked heterojunctions have garnered significant attention for their tunable electronic structures and photocatalytic performance, making them promising candidates for next-generation nanodevices. Using first-principles calculations, we systematically investigate the electronic structure, optical characteristics, and charge transfer of WSSe/SiC heterojunctions. Our results reveal that SeWS/monolayer-SiC, SeWS/bilayer-SiC, and SWSe/monolayer-SiC exhibit type-II band alignment, whereas SWSe/bilayer-SiC displays type-I alignment. Notably, SeWS/bilayer-SiC possesses a direct bandgap, in contrast to the indirect bandgaps of the other three configurations. Remarkably, the SeWS/bilayer-SiC heterojunction demonstrates a high absorption coefficient ($10^{5}~\mathrm{cm}^{-1}$) in the visible range and exhibits exceptional anisotropy in carrier transport, with an outstanding hole mobility of $9.58 \times 10^{3}~\mathrm{cm}^{2}\,\mathrm{V}^{-1}\,\mathrm{s}^{-1}$ along the Y-direction. Furthermore, combining thermodynamic stability with an S-scheme charge transfer mechanism, this system exhibits superior redox capability for photocatalytic water splitting, achieving a high hydrogen evolution efficiency of 22.15\%, which surpasses the commercial viability threshold (10\%). Furthermore, we demonstrate effective band gap modulation via external electric fields and biaxial strains, with optical absorption coefficients exhibiting strong strain dependence. This work provides fundamental insights into the design of WSSe/SiC heterojunctions for high-efficiency photocatalytic and tunable photodetector applications.
\end{abstract}

\maketitle


\maketitle
\section{INTRODUCTION}
Since the successful isolation of graphene in 2004 \cite{1a}, two-dimensional (2D) materials have rapidly emerged as a focal point in materials science \cite{2a,3a,4a,5a}. Graphene has attracted immense attention due to its exceptional physical and chemical properties, including high carrier mobility, superior mechanical strength, and remarkable thermal conductivity. Following this breakthrough, researchers have continuously explored a wide range of other 2D materials, such as silicon carbide (SiC)\cite{6a}, black phosphorus \cite{7a}, hexagonal boron nitride (h-BN) \cite{8a}, transition metal carbides and nitrides (MXenes)\cite{9a,10a,11a}, transition metal dichalcogenides (TMDs)\cite{12a,13a,14a,15a}, and so on. These 2D materials exhibit remarkable thickness-dependent properties in the electronic, optical, mechanical, and magnetic domains, that are highly desirable for next-generation nanoelectronic and optoelectronic applications. More importantly, van der Waals heterojunctions (vdWHs), formed by vertically stacking different 2D materials through weak vdW interactions, surpassing the performance limits of individual materials through interfacial coupling and band structure engineering, enabling novel quantum effects and multifunctional integration.

The vdWHs not only inherit the excellent properties of 2D monolayer materials but also exhibits some novel properties that monolayer materials do not possess and thus has received extensive attention in experimental and theoretical studies \cite{16a,17a,18a,19a,20a,21a}. The heterojunctions formed by two semiconductors are typically categorized based on the relative alignment of the valence band maximum (VBM) and conduction band minimum (CBM) of the constituent layers into three main types: type-I (straddling gap), type-II (staggered gap), and type-III (broken gap) band alignments. In type-I heterojunctions, both CBM and VBM reside within the same material, making them ideal for light-emitting applications\cite{22a,23a}. Type-II heterojunctions feature staggered band alignments, facilitating spatial separation of electrons and holes, which is beneficial for photovoltaic devices and photocatalytic water splitting\cite{24a,25a}. Type-III heterojunctions enable tunneling of charge carriers across the interface and are promising candidates for tunnel field-effect transistors (TFETs)\cite{26a,27a,28a}. Recent numerous theoretical and experimental studies have demonstrated the tunable optoelectronic characteristics of vdWHs such as SiC/SnSSe \cite{29a}, WSSe/g-GeC \cite{30a}, and WSSe/In$_2$Se$_3$ \cite{31a}. Studies have demonstrated that the interfacial properties of 2D heterojunctions play a significant role in modulating their electronic structures. Through theoretical calculations, Mogulkoc et al. \cite{32a} revealed that the interfacial coupling characteristics of boron phosphide (BP)/MoSSe heterojunction strongly depend on the contact orientation. When BP interacts with different surfaces of MoSSe, the resulting heterojunctions exhibit distinct interlayer spacings and bandgap values. Notably, under the regulation of external electric fields and biaxial strain, this system can achieve reversible transitions between type-I and type-II band alignments. On the other hand, the research group led by Zhang \cite{33a} fabricated InSe/PtTe$_2$ heterojunctions that demonstrate exceptional carrier transport properties, with hole mobility enhanced by two orders of magnitude compared to monolayer structures. Recently, Zhu et al. proposed that BiTeCl/GeSe vdWH is a type-S photocatalyst, which has a type-¢ò energy band arrangement, exhibits a high light absorption coefficient and a wide range of light absorption, and achieves a solar hydrogen production efficiency of up to 19.27\% \cite{34a}. These findings not only provide profound insights into interfacial effects in low-dimensional materials but also significantly advance the design and development of photocatalysts. The research results demonstrate that multidimensional regulatory factors (stacking configuration, interlayer spacing, strain, and electric field) can effectively modulate the properties of heterojunctions.

Among the 2D semiconductor family, Janus TMDs have attracted much attention due to their broken inversion symmetry and significant direct band gap within the visible light range \cite{35a}. In 2020, monolayer WSSe was successfully synthesized by pulsed laser deposition \cite{36a}. Meanwhile, in 2021, Chabi¡¯s group reported the top-down synthesis of monolayer SiC, a significant development considering that bulk SiC is not inherently layered\cite{37a}. Bulk SiC is a wide-bandgap semiconductor known for its exceptional thermal stability, high breakdown field, and strong mechanical strength \cite{38a}. When reduced to the 2D limit, SiC exhibits remarkable quantum confinement and surface effects, resulting in significantly altered optical and electrical properties \cite{39a}. Monolayer SiC has a wide band gap of more than 3.0 eV, and the carrier mobility shows obvious anisotropy. It has good light absorption in the ultraviolet region, with an absorption coefficient of $10^{5}~\mathrm{cm}^{-1}$. Yang et al. performed a comprehensive study on the structural, optical, and interlayer coupling properties of SiC nanosheets, revealing their promising characteristics for nano-optoelectronic applications\cite{40a}. Quantum confinement effects at the atomic level and interlayer interactions in multilayers cause the electrical and optical properties of 2D materials to change as the number of layers increases. Given that the properties of 2D materials are closely related to their number of layers, it is particularly important to probe deeply into how the number of layers affects the kinetic behavior of photoexcited carriers at heterojunctions interfaces \cite{41a,42a,43a,44a,45a}. Inspired by these studies, we constructed the WSSe/SiC vdWHs and systematically investigated its electronic structure, optical properties, and carrier mobility based on density functional theory. The results demonstrate that the WSSe/SiC vdWHs exhibits remarkable electrical and optical properties coupled with high carrier mobility, positioning it as a promising candidate material for future micro/nano-optoelectronic devices.

\section{ Computational Method}
In this study, we employ the Vienna Ab initio Simulation Package (VASP) \cite{46a,47a} to perform all density-functional theory (DFT) calculations. We have chosen the Perdew-Burke-Ernzerhof (PBE) generalized gradient approximation (GGA) \cite{48a} to describe the exchange-correlation energy. At the same time, the projection-enhanced wave (PAW) method \cite{49a} is utilized to deal with ion-electron interactions. To more accurately describe the van der Waals interactions between the layers, we also introduced the DFT-D3 method of Grimme with zero-damping function \cite{50a}. We set a truncation energy of 550 eV in the calculations and used a $12\times 12 \times 1$ K-point grid of Gamma points. The convergence criterion for the force was set to 0.001 eV/{\AA}, while the convergence criterion for the total energy was set to $10^{-6}$ eV to ensure the accuracy of the calculations. In order to eliminate artificial interactions that the periodic boundary conditions may introduce, we set up a vacuum layer of 20 {\AA} in the Z-direction. In addition, we employed the Heyd-Scuseria-Ernzerhof (HSE06) hybridization generalization \cite{51a} to calculate the band gap of the structure in order to obtain band gap values that are closer to the experimental ones. Considering the asymmetric structure of Janus WSSe, we also introduce dipole corrections to further enhance the accuracy of the calculation.

\section{Results and Discussion}
\subsection{ Crystal structure and stability}
In order to obtain accurate lattice parameters, careful geometrical optimization of monolayer-SiC(ML-SiC), bilayer-SiC(BL-SiC), and monolayer WSSe was first carried out. The optimization results are shown in Table 1 and are in high agreement with those reported in the existing literature \cite{40a,52a}. The optimized crystal structures are detailed in Fig. S1 in the Supplementary Material \cite{53a}. Due to the comparable lattice constants of SiC and WSSe, we constructed $2\times2\times1$ supercells for both constituents to form four types of vdWHs:  SeWS/monolayer-SiC (SeWS/ML-SiC), SeWS/bilayer-SiC (SeWS/BL-SiC), SWSe/monolayer-SiC (SWSe/ML-SiC), and SWSe/bilayer-SiC (SWSe/BL-SiC). For each heterojunction, six different stacking configurations were considered, labeled as $A1\sim A6, B1\sim B6, C1\sim C6$, and $D1\sim D6$, respectively, as illustrated in Fig. S2, to explore various interlayer atomic alignments and their effects on interfacial stability and electronic properties.
\begin{table}[h]
\begin{ruledtabular}
\caption{Optimized structural parameters of the monolithic material. $\textbf{a}$ denotes the lattice constant, $L_{\parallel}$ and $\theta^{\circ}_{\parallel}$ represent the in-plane bond length and bond angle, respectively, while $L_{\perp}$ and $\theta^{\circ}_{\perp}$ correspond to the out-of-plane bond length and bond angle. $E_{g}$ is the band gap(HSE06).}\label{tablep1}
\begin{tabular}{lccccccc}
Structure  & $\textbf{a}$ ({\AA}) & $L_{\parallel}$({\AA}) & $L_{\perp}$({\AA}) & $\theta^{\circ}_{\parallel}$ & $\theta^{\circ}_{\perp}$ & $E_{g}$(eV)   \\
   \hline

ML-SiC & 3.096 & 1.778 & ---    & 120.000 & ---                   & 3.36  \\

BL-SiC & 3.157 & 1.830 & 2.174  & 119.287 & 85.130$\sim$94.870    & 2.66  \\

WSSe   & 3.248 & --- & 2.423(W-S) &---    &79.498(Se-W-Se/W-Se-W) & 2.17   \\
       &  &    & 2.540(W-Se)      &       &81.708(S-W-Se)         & \\
       &  &    &                  &       &84.162(S-W-S/W-S-W)    & \\
\end{tabular}
\end{ruledtabular}
\end{table}

The structural stability of these heterojunctions was evaluated using three key criteria: lattice mismatch, formation energy, and phonon dispersion. The lattice mismatch was calculated using the following formulae:
\begin{equation}
L = (a_{1}-a_{2} )/a,
\end{equation}
 where ${a}_{1}, {a}_{2}$ are the lattice constants of WSSe and SiC, respectively, and ${a} = \frac{a_1 + a_2}{2}$ is the average value of the lattice constants of both. The computed lattice mismatches for the A1(C1) and B1(D1) stacking configurations are 4.79\% and 2.84\%, respectively. These values are within the acceptable range (typically less than 5\%) for constructing coherent and stable heterojunctions. To further assess the thermodynamic stability, the formation energy ($E_f$) was calculated as:
\begin{equation}
	E_f = E_H - E_{\mathrm{SiC}} - E_{\mathrm{WSSe}},
\end{equation}
where $E_H$, $E_{\mathrm{SiC}}$, and $E_{\mathrm{WSSe}}$ are the total energies of the heterojunction, isolated SiC, and WSSe, respectively. A negative $E_f$ signifies an energetically favorable, and thus thermodynamically stable structure. Among all the considered stacking arrangements, A1, B1, C1 and D1 have the lowest formation energies, which are -78.1 meV, -108.49 meV, -78.87 meV and -108.96 meV respectively (see Table S1). Therefore, we focus our subsequent analysis on these four most stable configurations, whose atomic structures are presented in Fig.~1(a-h). The equilibrium interlayer distances for the A1, B1, C1, and D1 configurations were found to be 3.041~\AA, 3.037~\AA, 3.680~\AA, and 3.586~\AA, respectively, within the typical range (3.0$\sim$3.8~\AA) for vdWHs. To better understand interfacial electronic interactions, we calculated the electron localization function (ELF)\cite{54a} for the four heterojunctions are (Fig. S4). The analysis reveals significant electron localization around the C, S, and Se atoms, and negligible electron localization is observed in the interfacial region between WSSe and SiC, confirming weak interlayer vdW interactions.

\begin{figure}[t!h]
\centerline{\includegraphics[width=1\textwidth]{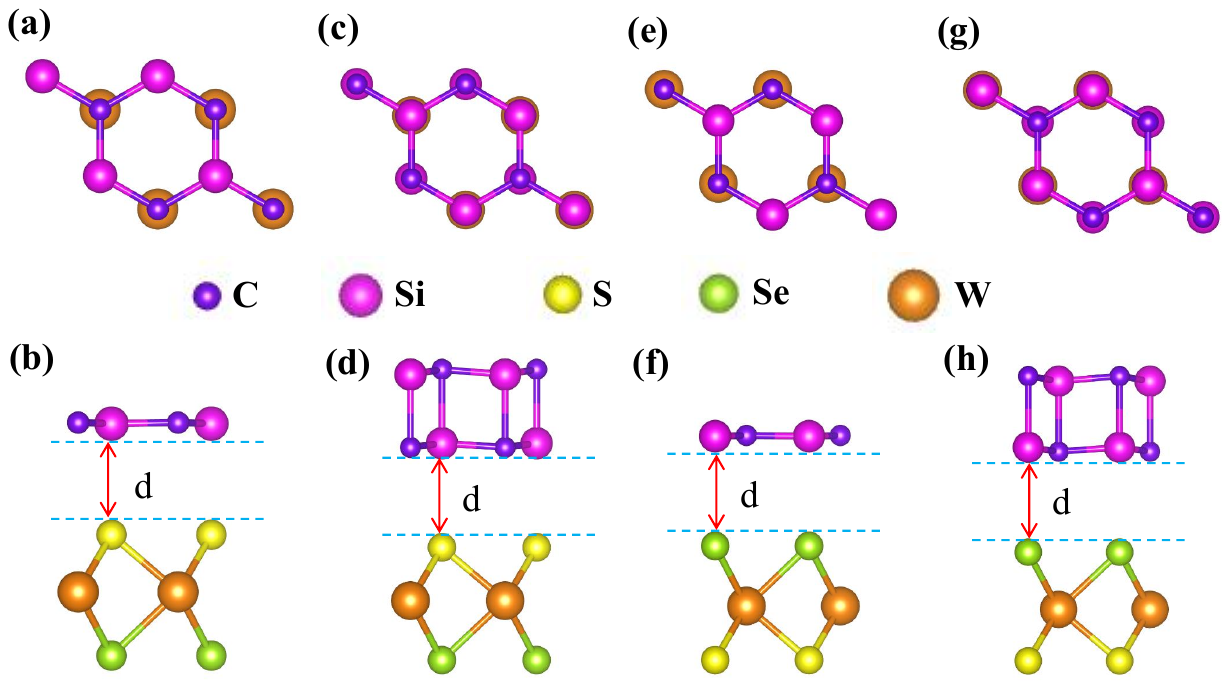}}
\vspace*{-40mm}
\caption{Top and side views of (a)(b) SeWS/ML-SiC, (c)(d) SeWS/BL-SiC, (e)(f) SWSe/ML-SiC, (g)(h) SWSe/BL-SiC after optimisation.
\label{fig:stru}}
\end{figure}

To validate the dynamical stability of the heterojunctions, we computed the phonon spectra for the A1, B1, C1, and D1 configurations. As shown in Fig. 2, none of the four selected configurations exhibit imaginary phonon frequencies, thereby affirming their dynamic stability. These results collectively indicate that the SeWS/ML-SiC, SeWS/BL-SiC, SWSe/ML-SiC, and SWSe/BL-SiC heterojunctions are structurally stable and are suitable for further investigation of their electronic and optical properties.
\begin{figure}[t!h]
\vspace*{-1mm}
\centerline{\includegraphics[width=1\textwidth]{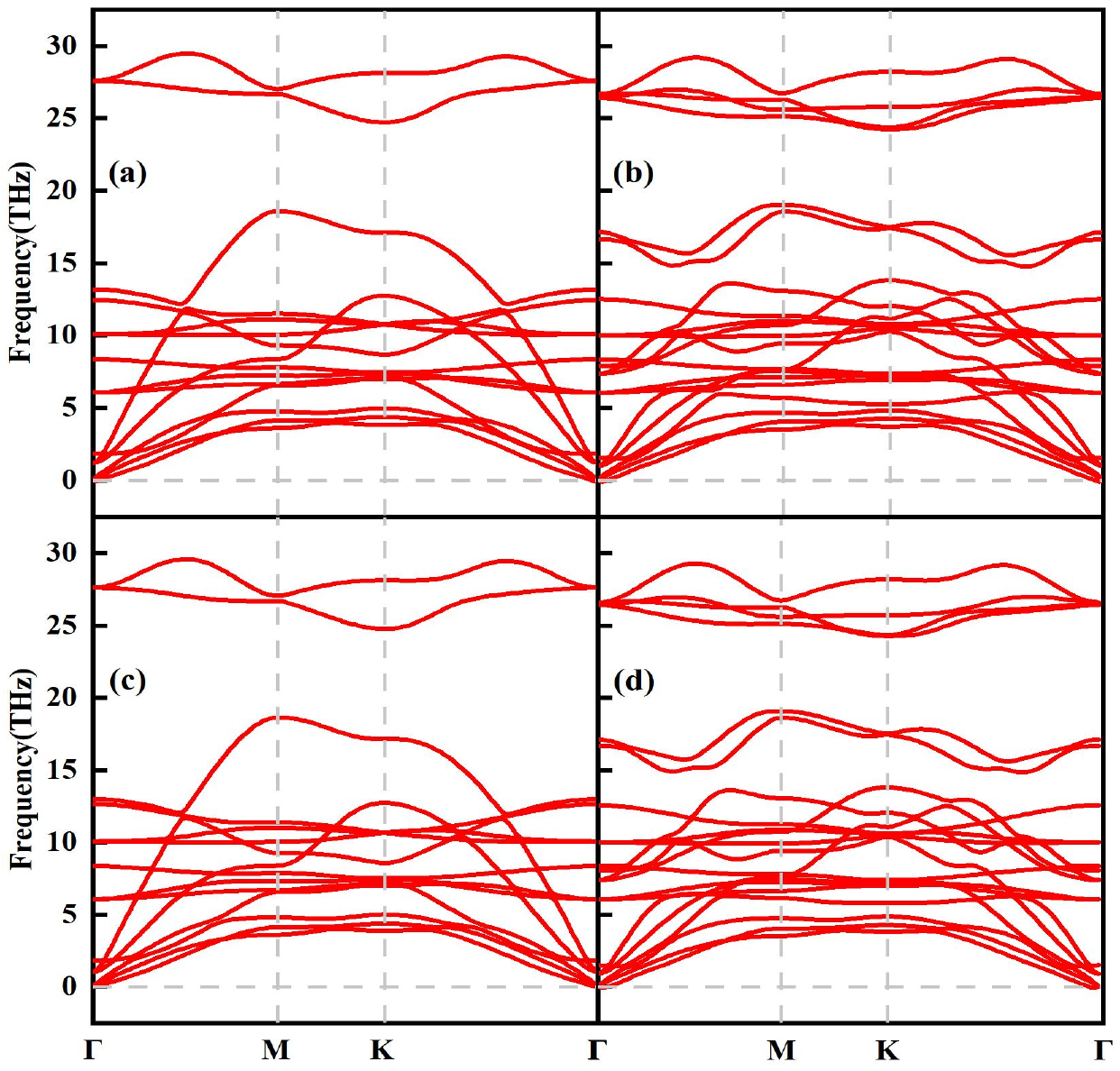}}
\vspace*{-1mm}
\caption{Phonon spectrum of (a) SeWS/ML-SiC, (b) SeWS/BL-SiC, (c) SWSe/ML-SiC, (d) SWSe/BL-SiC.
\label{fig:phonon}}
\end{figure}

\begin{figure}[t!h]
\vspace*{-1mm}
\centerline{\includegraphics[width=1\textwidth]{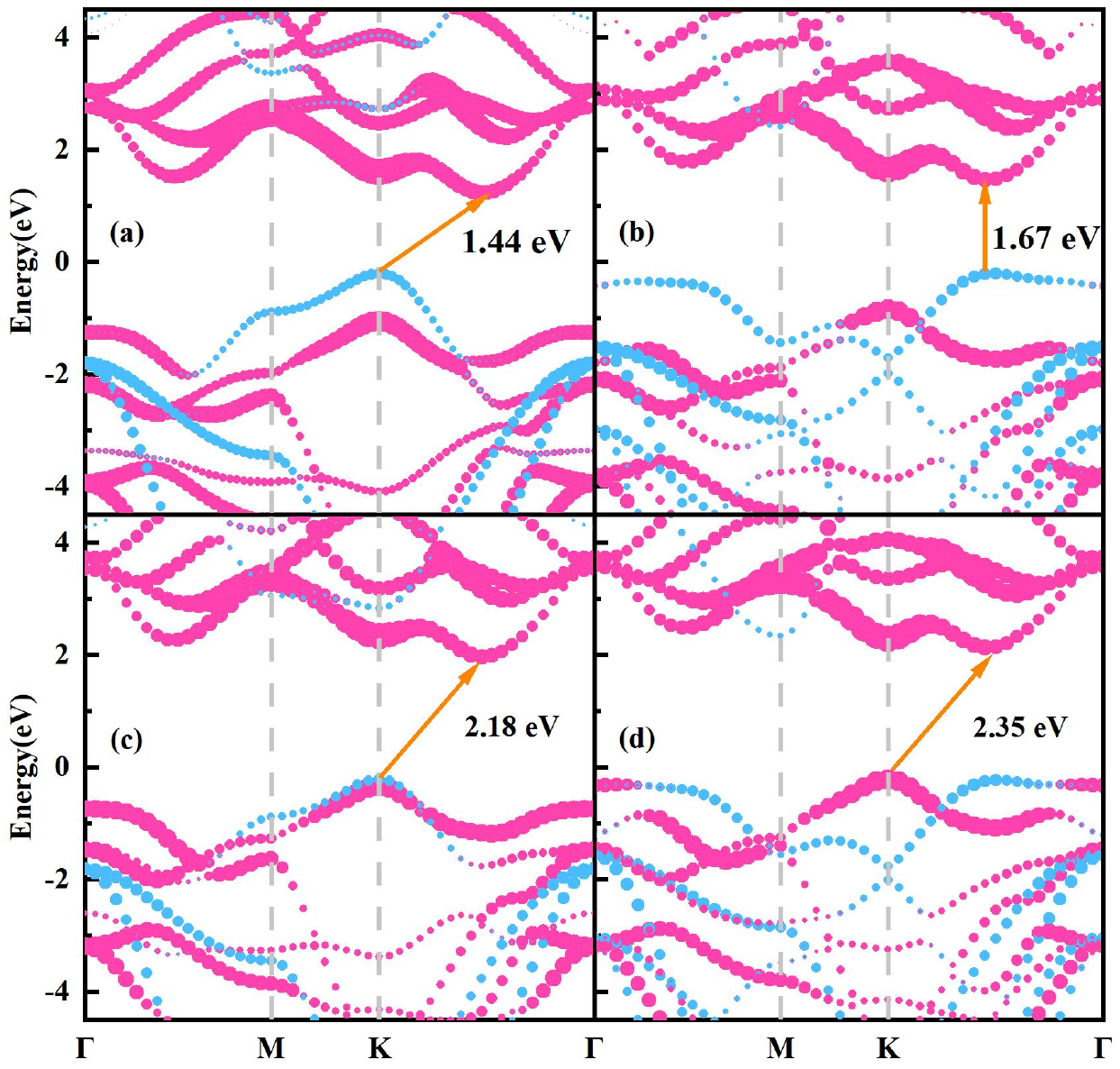}}
\vspace*{-1mm}
\caption{Projected energy band diagrams of (a) SeWS/ML-SiC, (b) SeWS/BL-SiC, (c) SWSe/ML-SiC, (d) SWSe/BL-SiC. The pink color is contributed by WSSe, and the blue color is contributed by SiC.
\label{fig:band}}
\end{figure}

\subsection{Electronic properties}
To explore the intrinsic electronic characteristics, we first calculated the band structures of ML-SiC, BL-SiC, and WSSe using HSE06 hybrid functional. As shown in Fig. S3, ML-SiC and BL-SiC are both indirect bandgap semiconductors, with bandgaps of 3.36 eV and 2.66 eV, respectively, agreement with previous theoretical studies\cite{6a,40a}. In contrast, WSSe has a direct bandgap of 2.17 eV. We further examined the projected energy band structures of the SeWS/ML-SiC, SeWS/BL-SiC, SWSe/ML-SiC, and SWSe/BL-SiC heterojunctions. As shown in Fig. 3(a), (b), and (c), the heterojunctions maintain semiconductor properties. The CBM is mainly contributed by WSSe, while the VBM is mainly derived from silicon carbide. This band edge distribution indicates that there is effective charge separation at the interface and it has the potential to form an intrinsic electric field. In contrast, in Fig. 3(d), both the CBM and VBM are contributed by WSSe.The SeWS/ML-SiC and SWSe/ML-SiC heterojunctions exhibit indirect band gaps of 1.44 eV and 2.18 eV, respectively, along with type-II band alignment. Interestingly, SeWS/BL-SiC heterojunction stands out with a direct bandgap of 1.67 eV and also exhibits type-II alignment, suggesting strong potential for photocatalytic water splitting. While, SWSe/BL-SiC heterojunction possesses an indirect band gap of 2.35 eV and a type-I (straddling) band alignment, where both the CBM and VBM reside in the WSSe. The heterojunctions thus not only reduces the wide band gap of pristine SiC but also extends its functional scope into diverse optoelectronic devices.

\begin{figure}[t!h]
\vspace{-1mm}
\centerline{\includegraphics[width=1\textwidth]{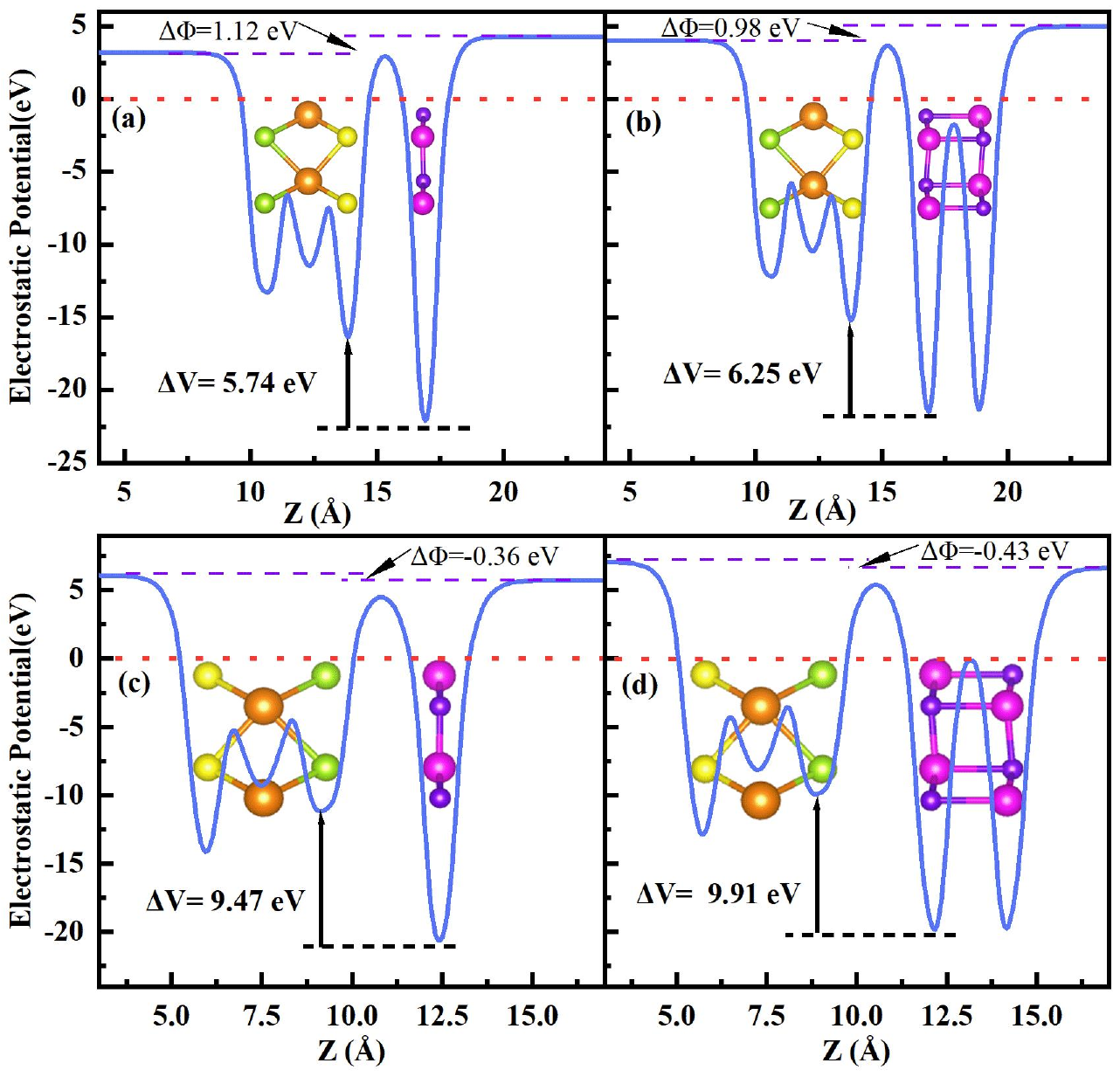}}
\vspace*{-1mm}
\caption{Electrostatic potential diagrams of (a) SeWS/ML-SiC, (b) SeWS/BL-SiC, (c) SWSe/ML-SiC, and (d) SWSe/ML-SiC. The red dashed line in the figure indicates the Fermi energy level, which is set to zero.
\label{fig:potential}}
\end{figure}

Furthermore, we carefully analyzed the electrostatic potential distribution and differential charge density characteristics of four heterojunctions to accurately quantify the dynamic process of charge transfer in the interface region. Fig. 4(a)$\sim$(d) present the electrostatic potential distributions of each heterojunction, respectively, from which we can observe that there exists a significant static potential difference ($\Delta$V) between the SiC layer and the WSSe layer, with the specific values of 6.20 eV, 6.29 eV, 9.48 eV, and 9.87 eV, respectively. This indicates that an electric field has been generated between silicon carbide and SWSe. Further by analyzing the differential charge density maps in Fig. 5(a) and (b), we clarified the exact path and magnitude of electron transfer: on the silicon carbide side, significant electron depletion was observed, while on the WSSe side, it was accompanied by significant electron accumulation. This electron migration phenomenon was strictly verified by Bader charge analysis, and the specific quantification is as follows: The charges transferred from the silicon carbide layer to the WSSe layer are 0.0829$|$e$|$ and 0.1156$|$e$|$ respectively. The electron transfer in Fig. 5(c) and (d) is the opposite, that is, the electrons transferred from the WSSe layer to the SiC layer are 0.0039$|$e$|$ and 0.0075$|$e$|$ respectively. We further investigated the work function ($W$)\cite{55a}(see Fig. 6), an essential quantity influencing charge transf. The work function can be calculated by the equation:
 \begin{equation}
 W = E_{vac} - E_{f},
 \end{equation}
  where $E_{vac}$ and $E_{f}$ represent the energy levels of the vacuum and Fermi, respectively. The difference in the work function between SiC and WSSe is a crucial factor contributing to the charge redistribution and the formation of the built-in electric field. Combining the work function, electrostatic potential and charge transfer, it can be concluded that the interface electric field direction of SeWS/ML-SiC and SeWS/BL-SiC is from the SiC layer to the SWSe layer, while that of SWSe/ML-SiC and SWSe/BL-SiC is the opposite. In particular, the SeWS/BL-SiC heterojunction is renowned for its unique Type II band arrangement and direct bandgap semiconductor characteristics.

 \begin{figure}[t!h]
\vspace*{-1mm}
\centerline{\includegraphics[width=1\textwidth]{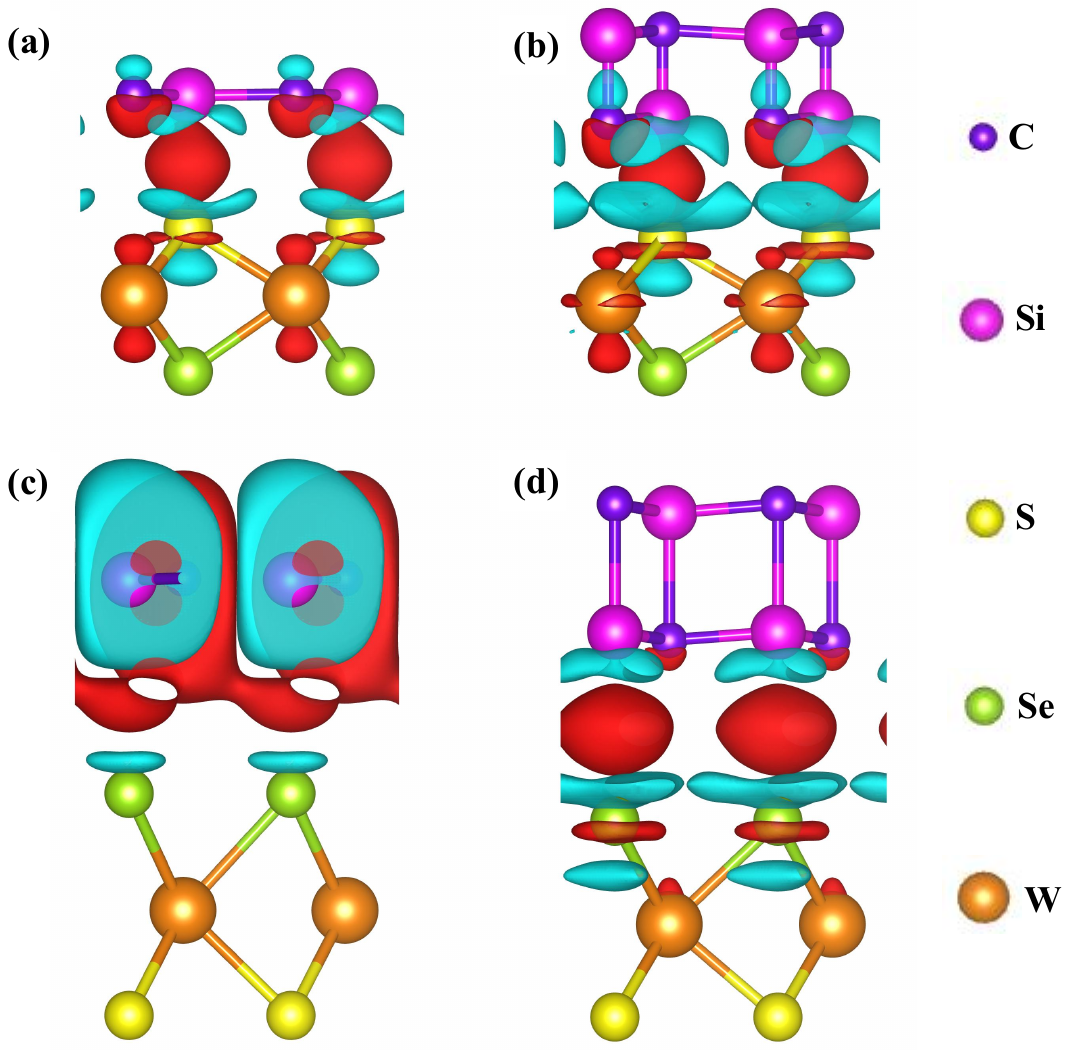}}
\vspace*{-1mm}
\caption{Differential charge density maps of (a) SeWS/ML-SiC, (b) SeWS/BL-SiC, (c) SWSe/ML-SiC, and (d) SWSe/ML-SiC. The red and cyan regions indicate the loss and accumulation of electrons, respectively. The iso-surface values of (a) and (b) are $3\times 10^{-3}$ e$\cdot${\AA}$^{-3}$, and the iso-surface values of (c) and (d) are $3\times 10^{-4}$ e$\cdot${\AA}$^{-3}$.
\label{fig:charge}}
\end{figure}

\begin{figure}[t!h]
\vspace*{-1mm}
\centerline{\includegraphics[width=1\textwidth]{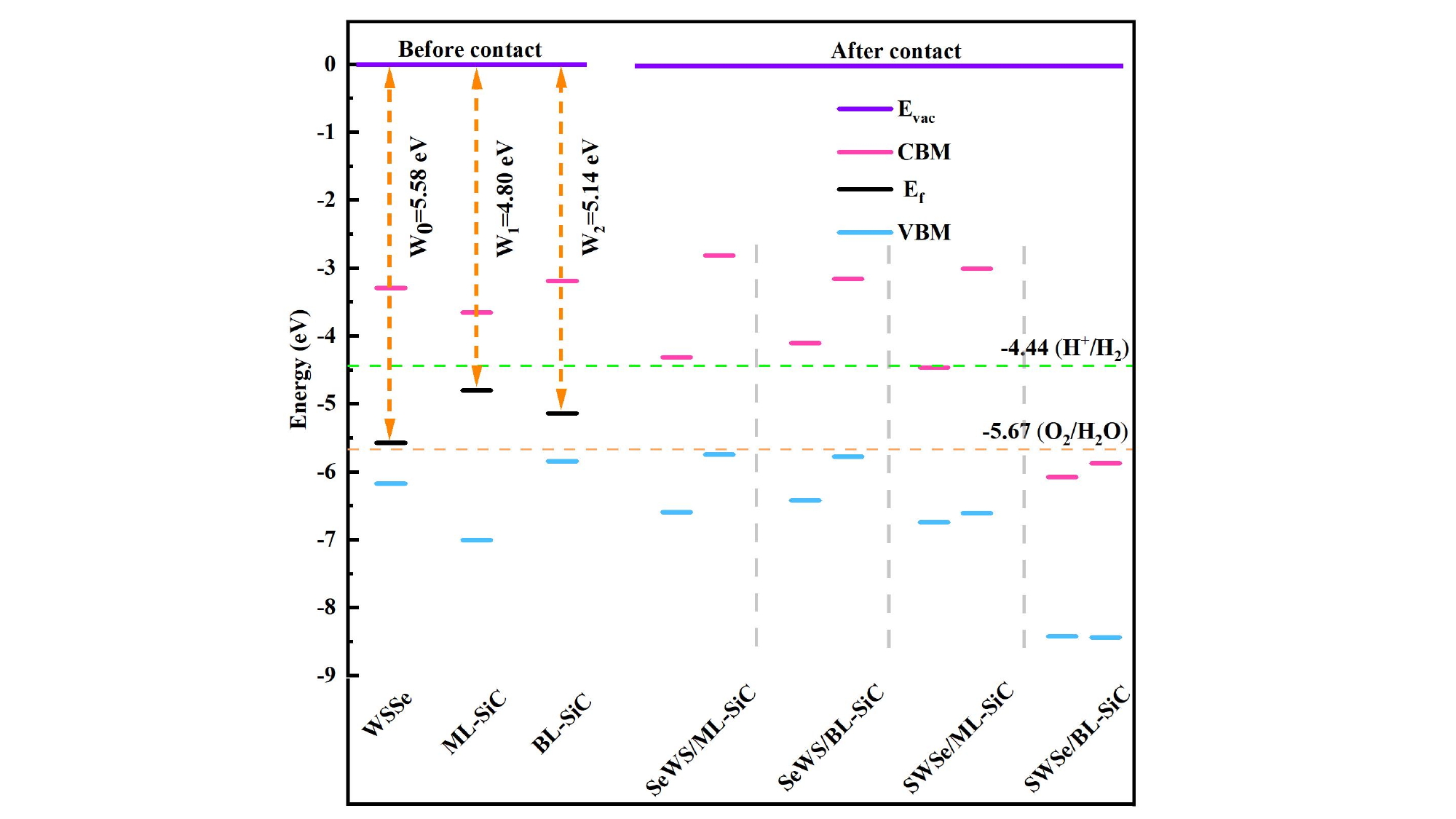}}
\vspace*{-1mm}
\caption{Changes in conduction band minimum (CBM), valence band maximum (VBM) of monolayer and heterojunctions before and after contacting. E$_{vac}$ is the vacuum energy level, set to 0. E$_{f}$ is the Fermi energy level. The orange and green dashed lines represent the oxidation and reduction potentials of water, respectively.
\label{fig:changes}}
\end{figure}

As previously illustrated in Fig. 6, BL-SiC has higher conduction band(CB) and valence band(VB) edge positions and a lower work function than SeWS. For the heterojunction, electrons spontaneously diffuse from BL-SiC to SeWS, generating an electron depletion zone on the BL-SiC side and an accumulation region on the SeWS side, reinforcing the built-in electric field pointing from BL-SiC to SeWS. Fig. 7(b) outlines the two primary charge transfer pathways occurring at the interface: (i) electrons transfer from the CB of BL-SiC to the CB of SeWS, while holes migrate from the VB of SeWS to the VB of BL-SiC; (ii) electrons in the CB of SeWS recombine with holes from the VB of BL-SiC at the interface. The built-in electric field from BL-SiC pointing towards SeWS accelerated the recombination process in the (ii) path and inhibited the recombination process in the (i) path (as shown in Fig. 7(b)). Furthermore, Fermi level alignment leads to an upward shift in the Fermi level of SeWS and a downward shift in that of BL-SiC, resulting in band bending. It promotes electron-hole recombination at the interface. Photogenerated electrons in the CB of SeWS and holes in the VB of BL-SiC are driven toward the interface by Coulombic attraction, facilitating recombination. In summary, three synergistic factors the interfacial electric field, band bending, and Coulombic attraction promote the recombination of low-energy charge carriers, while retaining high-energy electrons in the CB of BL-SiC and holes in the VB of SeWS for photocatalytic activity. This selective retention of high-energy carriers aligns with the principles of an S-scheme heterojunctions. Therefore, the SeWS/BL-SiC heterojunction can be classified as an S-scheme photocatalyst with significant potential for overall water splitting.

Given that heterojunction exhibit suitable band edge alignments for photocatalytic water splitting, we performed a detailed analysis of its interfacial charge transfer mechanism. The planar average charge density difference $\Delta \rho(z)$ for the SeWS/BL-SiC was calculated as:
\begin{equation}
	\Delta\rho(z) = \rho(z)_{\text{SeWS/BL-SiC}} - \rho(z)_{\text{SeWS}} - \rho(z)_{\text{BL-SiC}},
\end{equation}
where $\rho(z)_{\text{SeWS/BL-SiC}}$, $\rho(z)_{\text{SeWS}}$, and $\rho(z)_{\text{BL-SiC}}$ denote the planar average charge densities of the SeWS/BL-SiC heterojunction, isolated SeWS, and BL-SiC, respectively. The electron density primarily accumulates around the S atoms in the SeWS layer, resulting in a negatively charged SeWS surface and a corresponding positive charge on the BL-SiC, as shown in Fig. 7(a). This charge distribution generates an internal electric field oriented from the BL-SiC toward the SeWS layer. This is also consistent with the aforementioned analysis.

\begin{figure}[t!h]

\vspace*{-1mm}
\centerline{\includegraphics[width=1\textwidth]{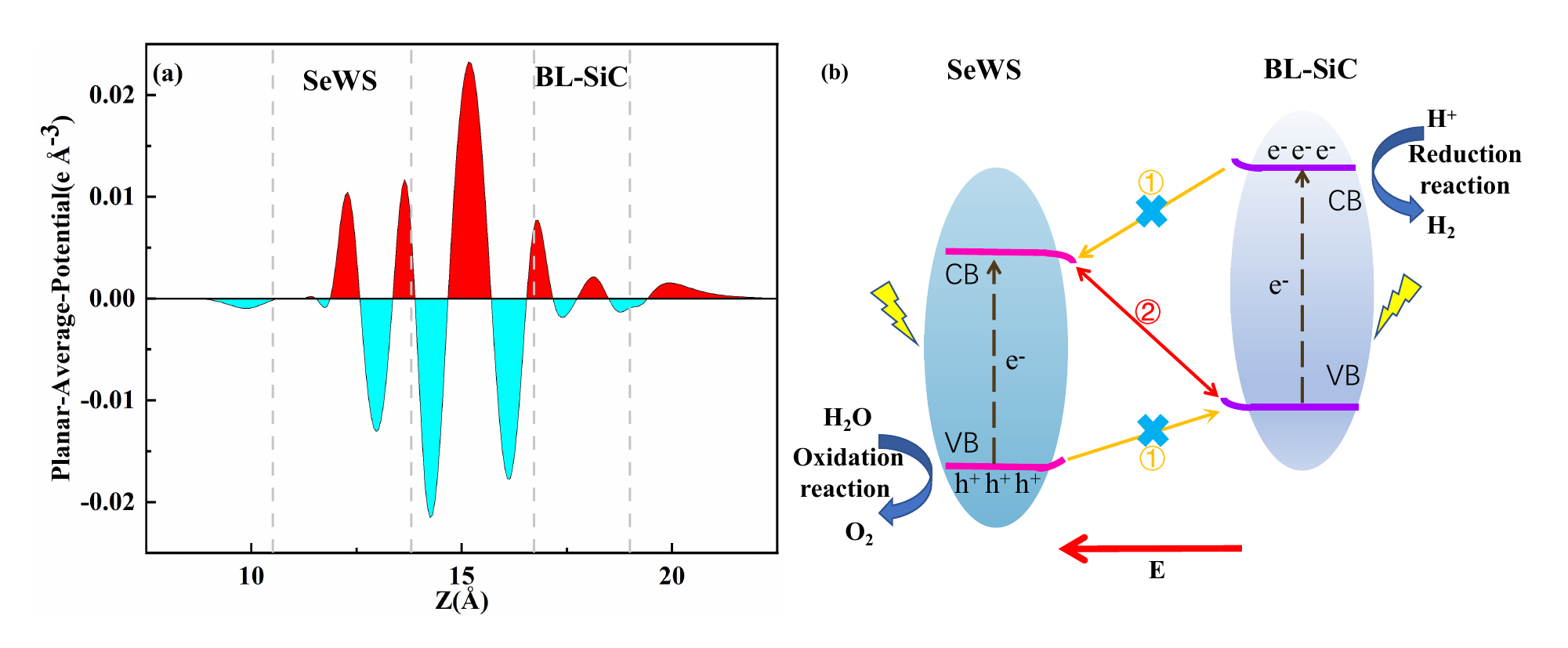}}
\vspace*{-1mm}
\caption{(a) the planar-averaged charge density difference of the SeWS/BL-SiC heterojunction. The red and cyan regions indicate the loss and accumulation of electrons, respectively. The dashed lines in the figure represent the interfacial boundaries of the materials.(b)Charge Transport Mechanism in SeWS/BL-SiC heterojunction.
\label{fig:planar}}
\end{figure}

\subsection{Carrier mobility and hydrogen production efficiency}
Carrier mobility is a key parameter in assessing the performance of photocatalytic and optoelectronic materials. While earlier studies commonly utilized the deformation potential theory by Bardeen and Shockley \cite{56a,57a}, it tends to overestimate carrier mobility in anisotropic materials due to its assumption of isotropic scattering \cite{58a,59a}. To address this limitation, Lang et al. derived an improved mobility expression for anisotropic 2D semiconductors using the Boltzmann transport equation under the relaxation time approximation.
The directional mobilities along the crystallographic axes are expressed as follows:

\begin{equation}
\mu_{x}=\frac{e \hbar^{3}\left(\frac{5 C_{11}+3 C_{22}}{8}\right)}{k_{\mathrm{B}} T\left(m_{x}\right)^{\frac{3}{2}}\left(m_{y}\right)^{\frac{1}{2}}\left(\frac{9 E_{1 x}^{2}+7 E_{1 x} E_{1 y}+4 E_{1 y}^{2}}{20}\right)},
\end{equation}
\begin{equation}
\mu_{y}=\frac{e \hbar^{3}\left(\frac{5 C_{22}+3 C_{11}}{8}\right)}{k_{\mathrm{B}} T\left(m_{y}\right)^{\frac{3}{2}}\left(m_{x}\right)^{\frac{1}{2}}\left(\frac{9 E_{1 y}^{2}+7 E_{1 y} E_{1 x}+4 E_{1 x}^{2}}{20}\right)},
\end{equation}

where $e$, $\hbar$, $k_B$, and $T$ represent the elementary charge, reduced Planck constant, Boltzmann constant, and absolute temperature (300 K), respectively. $m_x$ and $m_y$ are the effective masses of charge carriers along the $x$ and $y$ directions, $C_{11}$ and $C_{22}$ are the elastic constants, and $E_{1x}$, $E_{1y}$ are the deformation potential constants in the corresponding directions. To accurately reflect the in-plane anisotropy of the heterojunction, all mobility calculations were performed using an orthogonal unit cell instead of the conventional hexagonal lattice, as shown in Fig. S5. The calculated carrier mobilities are summarized in Table 2 and the heterojunction exhibit significant anisotropic transport behavior. Specifically, the hole mobility along the X-direction is approximately 11.7 times higher than the electron mobility, while in the Y-direction, the hole mobility is about 9.5 times that of electrons. Such anisotropy facilitates effective spatial separation of photogenerated carriers and suppresses charge recombination, thereby enhancing the photocatalytic efficiency for water splitting.

Theoretical calculations show that the SeWS/BL-SiC vdWH demonstrates three advantages: Firstly, it has a band edge position suitable for photocatalytic water decomposition and an ideal band gap (1.67 eV); Secondly, it shows excellent carrier mobility; Thirdly, it has a light absorption coefficient as high as $10^{5}~\mathrm{cm}^{-1}$ in the visible light band (1.59$\sim$3.26 eV) (see Fig. 9). In order to further evaluate its photocatalytic performance, the hydrogen production efficiency was calculated using the formula:
\begin{equation}
\eta_{S T H}=\eta_{a b s} \times \eta_{C u}
\end{equation}
where $\eta_{abs}$ and $\eta_{Cu}$ represent the light absorption efficiency and quantum conversion efficiency respectively.(for the specific calculation process, please refer to the supplementary materials). The results show that the light absorption efficiency of the SeWS/BL-SiC heterojunction reaches 51.16\%($\eta_{a b s}$), the quantum conversion efficiency is 52.59\%($\eta_{C u}$), and the original hydrogen production efficiency is 26.9\%($\eta_{S T H}$). Taking into account the influence of the potential difference ($\Delta\Phi$= 0.98 V, see Fig. 4(b)) caused by the built-in electric field, its hydrogen production efficiency is still as high as 22.15\%. This value significantly exceeds the reported WSSe/WSe$_{2}$(9.1\%) \cite{60a},Zr$_{2}$CO$_{2}$/WSe$_{2}$ (15.6\%) \cite{61a}, SiC/WS$_{2}$(16.91\%) \cite{62a}, and even breaks through the theoretical efficiency limit of traditional photocatalysis (18\%) \cite{63a}. The above research results indicate that the SeWS/BL-SiC vdWH is a promising photocatalyst. Therefore, in the following text, only further research will be conducted on SeWS/BL-SiC vdWH.

\begin{table}[h]
\begin{ruledtabular}
\caption{'e' and 'h' stand for 'electron' and 'hole' respectively. Elastic constants $C_{11}$ and $C_{22}$$(\mathrm{N  m^{-1}})$. Linear fitting of the positions of CBM and VBM in the strain range of -2\% to 2\% gives the deformation potential $\mathrm{E_{lx}}$ and $\mathrm{E_{ly}}$(eV), effective mass $m_{x}$ and $m_{y}$ are in unit of $m_{0}=9.11\times10^{-31}$ kg, $\mu_{x}$ and $\mu_{y}$ are expressed in units of $\mathrm{10^{3}  cm^{2}V^{-1}s^{-1}}$. $\mathrm{E_{lx}}$, $m_{x}$, and $\mu_{x}$ denote the X-direction, while $\mathrm{E_{ly}}$, $m_{y}$, and $\mu_{y}$ represent the Y-direction.}\label{tablep1}
\begin{tabular}{lccccccccc}
Structure  & Carrier type & $C_{11}$ & $C_{22}$ & $\mathrm{E_{lx}}$ & $\mathrm{E_{ly}}$ & $m_{x}$ & $m_{y}$ & $\mu_{x}$ & $\mu_{y}$ \\
   \hline

SeWS/BL-SiC & e & 365.430 & 365.625    & 4.280 & 5.286 & 0.862 & 0.626 & 0.565 & 1.010 \\
            & h & 365.430 & 365.625    & 1.582 & 1.604 & 0.750 & 0.514 & 6.610 & 9.580 \\

\end{tabular}
\end{ruledtabular}
\end{table}

\subsection{Electric field and strain modulation}
In practical electronic and optoelectronic applications, gate voltage is frequently employed to modulate the charge carrier dynamics and overall performance of semiconductor. This study investigates the effects of vertical electric fields on electronic properties. It is also noteworthy that the direction of the positive external electric field is along the positive direction of the Z-axis, which is opposite to the inherent electric field direction within the semiconductor. As shown in Fig.~8, the CBM of SeWS/BL-SiC vdWH exhibits a continuous upward shift within the applied electric field range of $-0.5$ V/\AA{} to $0.5$ V/\AA{}, whereas the VBM remains relatively unchanged. Consequently, the bandgap exhibits a monotonic increase with the strengthening of external electric field.

As is well known, strain engineering has emerged as an effective strategy to tune the electronic structure and improve the performance of 2D materials\cite{64a}. In this study, we systematically examined the effect of biaxial strain on the SeWS/BL-SiC vdWH by modifying the in-plane lattice constants and analyzed the changes in CBM, VBM, and bandgap. The following equation calculated the strain $\varepsilon$:
\begin{equation}
\varepsilon=\frac{a-a_{0}}{a_{0}} \times 100 \%,
\end{equation}
where $a$ and $a_0$ represent the strained and unstrained lattice constants, respectively. The strain-dependent variation of the CBM, VBM, and band gap is shown in Fig.~8(b). A monotonic bandgap narrowing is observed under tensile strain in the 0\% to 6\% range, whereas compressive strain induces an initial bandgap widening followed by narrowing. Additionally, Fig. S7 reveals that strain can induce a direct-to-indirect bandgap transition.

\begin{figure}[t!h]
\vspace*{-1mm}
\centerline{\includegraphics[width=1\textwidth]{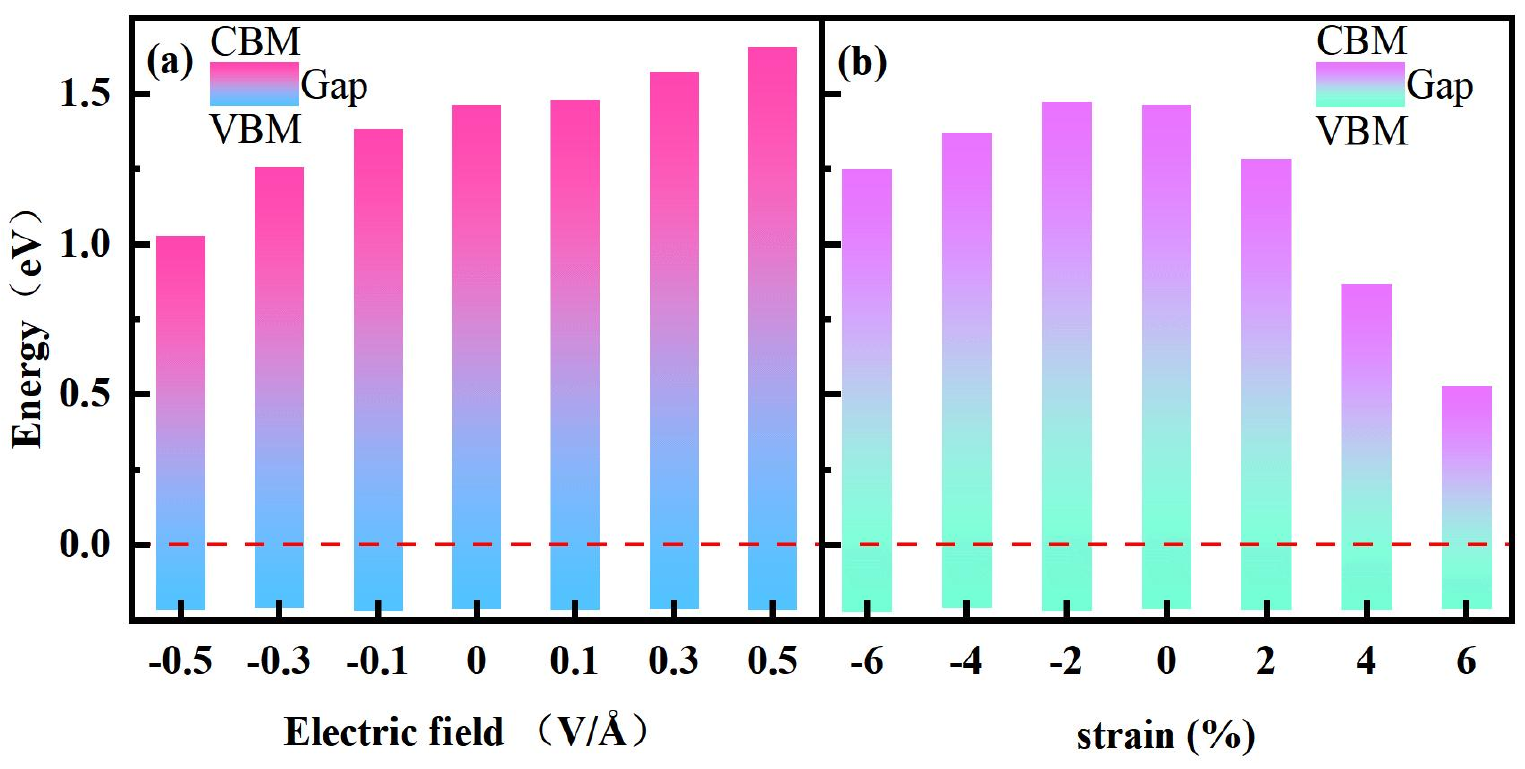}}
\vspace*{-40mm}
\caption{(a) CBM, VBM, and band gap trends of SeWS/BL-SiC under applied electric field. (b) CBM, VBM, and band gap trends of SeWS/BL-SiC under biaxial strain. The upper boundary of the data column indicates the CBM, the lower boundary indicates the VBM and the height indicates the Gap value. The red dashed line in the figure indicates the Fermi energy level, which is set to zero.
\label{fig:field}}
\end{figure}

\subsection{Optical properties}
The optical response of a heterojunction can be evaluated by analyzing its absorption coefficient $\alpha(\omega)$, which can be computed using the following relation:
\begin{equation}
	\alpha(\omega) = \sqrt{2}\omega \left\{ \left[\varepsilon_1^2(\omega) + \varepsilon_2^2(\omega)\right]^{\frac{1}{2}} - \varepsilon_1(\omega) \right\}^{\frac{1}{2}},
\end{equation}
where $\varepsilon_1(\omega)$ and $\varepsilon_2(\omega)$ are the real and imaginary parts of the complex dielectric function, respectively.

\begin{figure}[t!h]
\vspace*{-1mm}
\centerline{\includegraphics[width=1\textwidth]{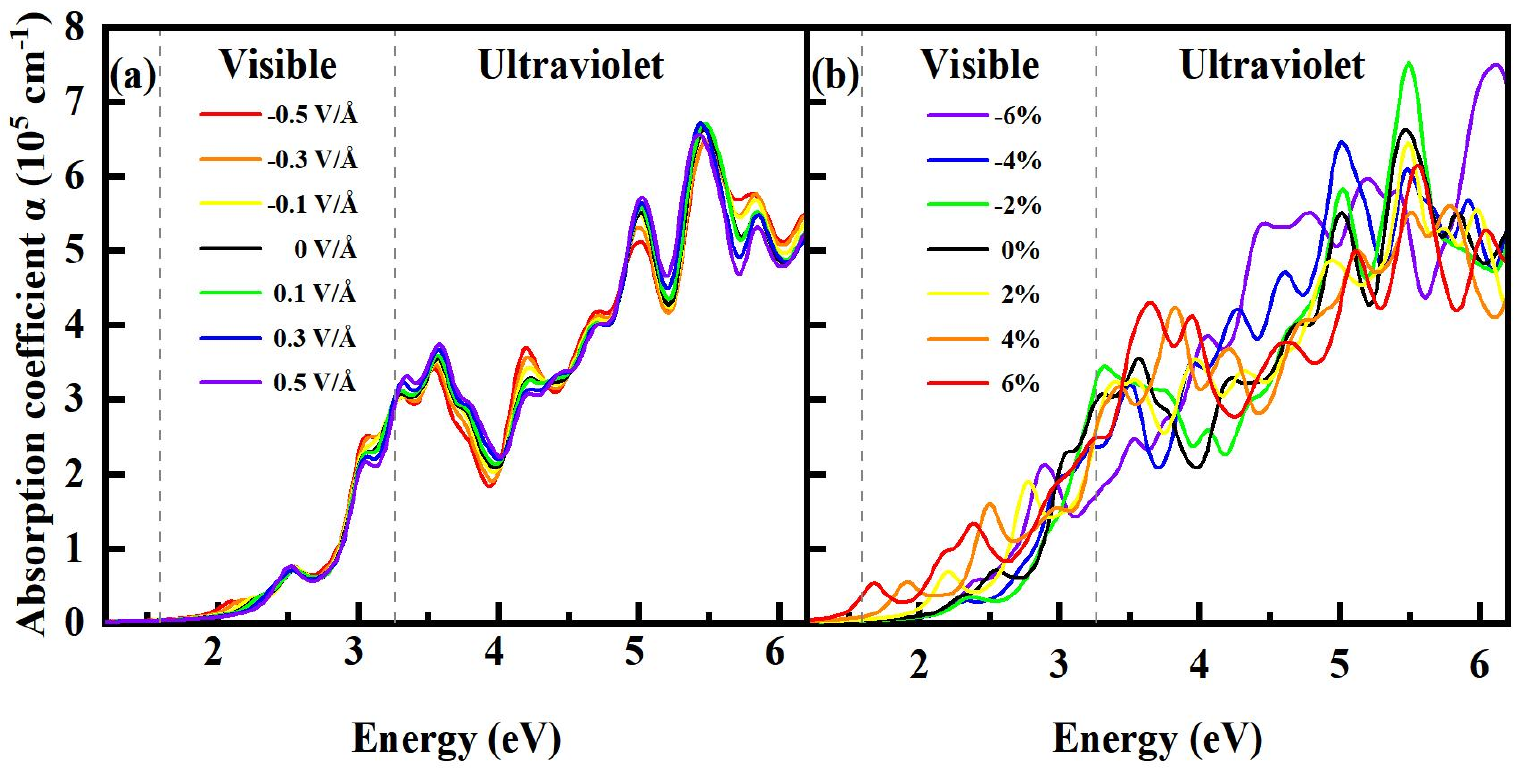}}
\vspace*{-40mm}
\caption{(a) Absorption spectra of SeWS/BL-SiC under applied electric field; (b) Absorption spectra of SeWS/BL-SiC under biaxial strain.
\label{fig:absorption}}
\end{figure}

Fig. 9 shows the variations of the absorption spectra under different external electric fields and biaxial strains. In the visible light range, under a reverse electric field($Z<0$), the absorption edge exhibits a red shift, and the absorption peak increases in the visible region. A notable absorption peak appears near 2.2 eV (approximately 560 nm). Conversely, under a forward electric field($Z>0$), the absorption edge blue-shifts and the peak intensity weakens, consistent with the field-induced widening of the band gap (Fig. S6). For the ultraviolet (UV) region, the behavior is reversed: the forward electric field($Z>0$) enhances absorption, while the reverse field suppresses it, with peak values reaching up to $6.5 \times 10^5$ cm$^{-1}$. These observations confirm that the SeWS/BL-SiC vdWH exhibits strong light absorption across both visible and UV regions. Moreover, a distinct red shift of the absorption edge is observed under tensile strain ($0\%$ to $6\%$), accompanied by significant enhancement in the visible light absorption. This enhancement arises from the bandgap narrowing. And the compressive strain enhances the UV absorption, with the highest absorption coefficient reaching $7.5 \times 10^{5}$~cm$^{-1}$ at $-2\%$ strain.

Our findings reveal the remarkable potential of SeWS/BL-SiC vdWH for diverse optoelectronic applications, including photovoltaic cells, high-performance electronic devices (particularly switching components), tunable multicolor LEDs, photocatalysis, and optical sensors. The system's electronic band structure demonstrates exceptional tunability through both electric field gating and biaxial strain engineering, enabling precise control over charge transport properties (conductivity), emission characteristics (LED wavelengths), and spectral absorption profiles (photocatalytic activity). Notably, strain-mediated bandgap narrowing and the concomitant redshift of absorption peaks significantly enhance visible-light harvesting efficiency, which simultaneously boosts photocatalytic performance and improves sensor responsiveness. These multifunctional capabilities position SeWS/BL-SiC vdWH as a promising platform for next-generation technologies in renewable energy conversion, environmental monitoring systems, and smart industrial automation.

\section{Conclusion}
In summary, through comprehensive first-principles calculations, we have systematically investigated the structural, electronic, and optical properties of four distinct WSSe/SiC vdWHs. Band structure analysis reveals the following characteristics for each system: (i) SeWS/ML-SiC forms a type-II indirect bandgap semiconductor (1.44 eV), (ii) SWSe/ML-SiC exhibits a type-II indirect bandgap (2.18 eV), (iii) SWSe/BL-SiC displays a type-I indirect bandgap (2.35 eV), while (iv) SeWS/BL-SiC emerges as a type-II direct bandgap material (1.67 eV). Notably, the SeWS/BL-SiC heterojunction establishes an S-scheme charge transfer pathway, demonstrating exceptional redox capability for photocatalytic applications. Further analysis reveals significant anisotropy in charge carrier transport properties, with hole mobility reaching an outstanding value of $9.58 \times 10^{3}~\mathrm{cm}^{2}\,\mathrm{V}^{-1}\,\mathrm{s}^{-1}$ along the Y-direction. The SeWS/BL-SiC vdWH also exhibits remarkable visible-light absorption characteristics, with absorption coefficients as high as $10^{5}~\mathrm{cm}^{-1}$. External field modulation studies indicate that the bandgap of SeWS/BL-SiC can be systematically tuned by applying an electric field (-0.5 V/{\AA} to 0.5 V/{\AA}), while both negative electric fields and tensile strain significantly enhance its visible-light absorption capacity. Most remarkably, the SeWS/BL-SiC heterojunction achieves an impressive hydrogen evolution efficiency of 22.15\%, highlighting its potential as a high-performance photocatalyst. These findings collectively demonstrate that the SeWS/BL-SiC vdWH holds great promise for a wide range of applications, including photovoltaic devices, photocatalytic water-splitting, optoelectronic components, and tunable functional devices.

\section{Acknowledgement}
This work was supported by the Innovation Capability Improvement Project of Hebei province (No. 22567605H). The numerical calculations in this paper have been done on the supercomputing system in the High Performance Computing Center of Yanshan University.


\end{document}